\documentclass[a4paper,11pt]{article}
\usepackage{pos}
\usepackage{subcaption}

\newcommand{\eps}{\varepsilon}

\newcommand{\Fgeom}{F_{\mathrm{geom}}}

\title{The geometric bookkeeping guide for $\varepsilon$-factorised differential equations}
\author*[a]{Antonela Matija\v{s}i\'{c}}

\affiliation[a]{PRISMA Cluster of Excellence, Institut f{\"u}r Physik, Johannes Gutenberg-Universit\"at Mainz,\\ D-55099 Mainz, Germany, }

\emailAdd{amatijas@uni-mainz.de}

\abstract{Precision predictions for high-energy experiments rely on accurately evaluating multi-loop, multi‑scale Feynman integrals in dimensional regularisation. The method of differential equations is by now the standard tool for this task, but its full power is only realised when the system can be brought into an $\varepsilon$-factorised form. In this talk, we present an algorithmic framework that systematically constructs $\varepsilon$-factorised differential equations for arbitrary integral families, independent of their underlying geometry.

We work in the setting of twisted cohomology and study the space of differential forms associated with a given family of Feynman integrals in the Baikov representation. Our approach consists of two steps. First, we introduce a particular ordering for the Laporta algorithm that orders Feynman integrals within a sector according to their geometric properties. We observe that this order relation yields a basis whose differential equation is in a Laurent polynomial form in the dimensional regulator $\varepsilon$. In the second step, we systematically construct transformation matrices such that the resulting system is in the $\varepsilon$‑factorised form. }

\FullConference{Loops and Legs in Quantum Field Theory (LL2026)\\
12-17, April, 2026\\
Bayreuth, Germany\\}

\begin{document}
\maketitle

\section{Introduction}

In recent years, progress in high-energy physics has been largely driven by the availability of precision calculations, which depend extensively on perturbative quantum field theory and Feynman integrals. These computations are essential not only for high-energy physics itself, but also play an important role in gravitational wave physics and in high-precision experiments at low energies. One of the main bottlenecks in these computations is efficiently evaluating multi-loop and multi-scale Feynman integrals.

By now, the standard tool for evaluating Feynman integrals is the method of differential equations~\cite{Kotikov:1991pm,Remiddi:1997ny,Gehrmann:1999as}. In the first step, a finite basis of master integrals is determined, along with the differential equations they satisfy using integration-by-parts identities~\cite{Tkachov:1981wb, Chetyrkin:1981qh} and the Laporta algorithm~\cite{Laporta:2000dsw}. The resulting differential equations can be solved either analytically or numerically~\cite{Liu:2017jxz,Liu:2022chg,Liu:2022mfb,Hidding:2020ytt,Armadillo:2022ugh,Prisco:2025wqs, PetitRosas:2025xhm}. In analytic computations, one typically first identifies a transformation to an $\eps$-factorised system of differential equations~\cite{Henn:2013pwa}, which can then be solved order-by-order in $\eps$ using iterated integrals~\cite{Chen:1977oja}.

There are a few bottlenecks with this approach. First, despite the existence of several optimised public implementations of integration-by-parts reduction~\cite{Peraro:2019svx,Lee:2013mka,Guan:2024byi,Lange:2025fba,Smirnov:2025prc}, in state-of-the-art applications one can still encounter limitations due to available computing resources. Second challenge comes from finding a transformation to $\eps$-factorised differential equation especially in presence of non-trivial geometries (see~\cite{Bree:2025tug,Forner:2026vby} and references therein).

In~\cite{e-collaboration:2025frv, Bree:2025tug}, these challenges are addressed by a two-step procedure for finding $\eps$-factorised differential equations. In the initial step, we use a new ordering relation in the Laporta algorithm which avoids spurious polynomials in the reduction coefficients. We find that this ordering relation leads to a basis whose differential equation is expressed as a Laurent polynomial in $\varepsilon$. Furthermore, this differential equation can be systematically transformed into an $\varepsilon$‑factorised form without using the knowledge of the underlying geometry.

In this talk, we review the main steps of the proposed algorithm and demonstrate its application on an example relevant for low-energy physics precision computations.

\section{Setup and algorithm overview}
We are interested in a family of Feynman integrals and we assume that any scalar product involving loop momenta can be written as a linear combination of inverse propagators and a constant term, i.e. we introduce irreducible scalar products (ISPs) if needed. 

We would like to improve the ordering within a given sector, therefore we can work on the maximal cut. It is convenient to analyse the maximal cut of a Feynman integral in Baikov representation~\cite{Baikov:1996iu}. In particular, we are using a loop-by-loop Baikov representation~\cite{Frellesvig:2017aai} 
\begin{equation}
    I_{\nu_1,\ldots,\nu_{N_E}}\left(D_{int},\varepsilon,x\right) = C_{B} \int \prod_{i\in I_{all}}[p_i(z,x)]^{\alpha_i} \prod_{j=1}^{N_E} \dfrac{d z_j}{z_j^{\nu_j}},
\end{equation}
where $z_j$ are Baikov variables, and $C_B$ depends on kinematic variables $x$ and dimensional regulator $\eps$. Exponents $\alpha_i$ are always of the following form
\begin{equation}
    \alpha_i=\frac{1}{2}\left(a_i + b_i \varepsilon \right), \quad a_i,\;b_i \in \mathbb{Z}.
\end{equation}
We say that polynomial $p_i$ is even if the corresponding $a_i$ is even. Similarly, if $a_i$ is odd, we call the corresponding polynomial $p_i$ odd.

On the maximal cut, we have
\begin{equation}
     C_{B} \int_{\mathcal{C}_{\text{maxcut}}} \dfrac{d^n z}{(2\pi i)^n} \prod_{i\in I_{all}}[p_i(z,x)]^{\alpha_i},
\end{equation} 
where $z$ are now remaining $n$ Baikov variables. We extend the affine space with coordinates $(z_1,\ldots,z_n)$ to projective space with homogeneous coordinates $[z_0:z_1:\ldots:z_n]$. Furthermore, we define differential forms as
\begin{equation}
    \label{eq:diff_forms}
    \Psi_{\mu_0,\ldots, \mu_{N_D}}[Q]=C\;U\; \dfrac{Q}{\prod_{i\in I_{all}^0} P_i^{\mu_i}} \eta, \quad \mu_i \in \mathbb{N}_0.
\end{equation}
Here, $C$ is $z$-independent prefactor (see Ref.~\cite{e-collaboration:2025frv, Bree:2025tug} for definition), $U$ is a twist function
\begin{equation}
    U=\prod_{i\in I_{odd}}[P_i(z,x)]^{-\frac{1}{2}+\frac{1}{2} b_i \varepsilon}\prod_{i\in I_{even}}[P_i(z,x)]^{\frac{1}{2} b_i \varepsilon},
\end{equation}
$Q$ is a homogeneous polynomial in Baikov variables, and $\eta$ is the standard differential $n$-form.

These differential forms satisfy three types of linear relations:
\begin{enumerate}
    \item Integration-by-parts identities
    \begin{equation}
        \frac{1}{\varepsilon}\Psi_{\mu_0\ldots\mu_i\ldots\mu_{N_D}}\left[\partial_{z_j}Q_{+}\right] + \sum_{i\in I_{all}} \Psi_{\mu_0\ldots(\mu_i+1)\ldots\mu_{N_D}}\left[Q_{+}\left(\partial_{z_j} P_i \right)\right] =0
    \end{equation}
    \item Distribution identities
    \begin{equation}
        \Psi_{\mu_0\ldots\mu_{N_D}}\left[Q_1 + Q_2\right] = \Psi_{\mu_0\ldots\mu_{N_D}}\left[Q_1 \right]+\Psi_{\mu_0\ldots\mu_{N_D}}\left[Q_2\right]
    \end{equation}
    \item Cancellation identities
    \begin{equation}
        \Psi_{\mu_0\ldots(\mu_j+1)\ldots \mu_{N_D}}\left[P_j Q\right] = \frac{1}{\varepsilon} \left(\frac{a_j}{2}-\mu_j+\frac{b_j}{2}\varepsilon\right)\Psi_{\mu_0\ldots\mu_j\ldots \mu_{N_D}}\left[Q\right].
        \label{eq:ci}
    \end{equation}
\end{enumerate}
After modding out these linear relations, differential forms span a finite-dimensional vector space $H^n_{\omega}$ (twisted cohomology group).

We would like to minimise the division with the linear factors in $\eps$ on the right hand side of the cancellation identities~\eqref{eq:ci}. Therefore, within a given sector, we propose a new ordering criteria
\begin{equation}
    \label{eq:new ordering}
    \left(a,w,o,|\mu|,\ldots \right).
\end{equation}
First we order the differential forms by the localization level $a$. If even polynomials appear in the denominator of a differential form, we may take a residue around $P_i=0, \; i \in I_{even}^0$. We then say that we localise at this specific $P_i$, which yields a simpler problem. This procedure should be iterated until we get to a point. By assigning $a=-w$ to preferred candidates that are masters on a localization and $a=0$ to all other forms,
%\begin{align}
%     a = \left\{
% \begin{array}{rl}
% -w, & \Psi_{\mu_0 \dots \mu_{N_D}}[Q] \; \mbox{is a preferred candidate on a localisation}\\
% 0, & \mbox{otherwise,} \\
% \end{array}
% \right.
%\end{align}
we want to give preference to differential forms coming from these simpler problems. The next integer we assign to a differential forms is the Hodge weight $w=n+r$, where $r$ is the number of consecutive non-zero residues that we can take. Within a given weight, we organise differential forms by pole order $o$. Both $w$ and $o$ are computed after setting $\eps = 0$ in $U$. Finally, the last integer we assign is $|\mu|=\sum_{i \in I_{all}} \mu_i$.

In addition to being used in the ordering relation, integers $(r,o,|\mu|)$ define three filtrations $W_{\bullet}, \; F_{geom}^{\bullet},\; F_{comb}^{\bullet}$, respectively, on the space of differential forms. These filtrations can be used to decompose the vector space of differential forms into subspaces.

Using the ordering criteria~\eqref{eq:new ordering} in the Laporta algorithm we obtain a basis $J$.
We observe that the coefficients in the reduction of the derivatives of master integrals $J$ are Laurent polynomials in $\eps$, i.e. we get a differential equation of the following form
\begin{equation}
    \text{d} J = \sum_{k=k_{min}}^1 \varepsilon^k A^{(k)}(x) J.
\end{equation}
In addition, if $\Psi_i$ has $|\mu|=|\mu|_i$ and $\Psi_j$ has $|\mu|=|\mu|_j$, we observe that 
\begin{equation}
    A_{i,j}(\eps,x)=\sum_{k_{min}=-|\mu|_i-|\mu|_j}  \eps^k A_{i,j}^{(k)}(x),
\end{equation}
and we call such a differential equation $F^{\bullet}$-compatible differential equation with respect to $F_{comb}^{\bullet}$ filtration.

In Ref.~\cite{Bree:2025tug}, we prove that $F^{\bullet}$-compatible differential equation can always be systematically transformed to an $\eps$-factorised differential equation. 

\section{Illustrative example: a kite diagram}
Let us take a look at a particular example and go through the main steps of the algorithm.

\begin{figure}[t!]
    \begin{subfigure}{0.5\textwidth}
    \centering
    \includegraphics[width=0.6\linewidth]{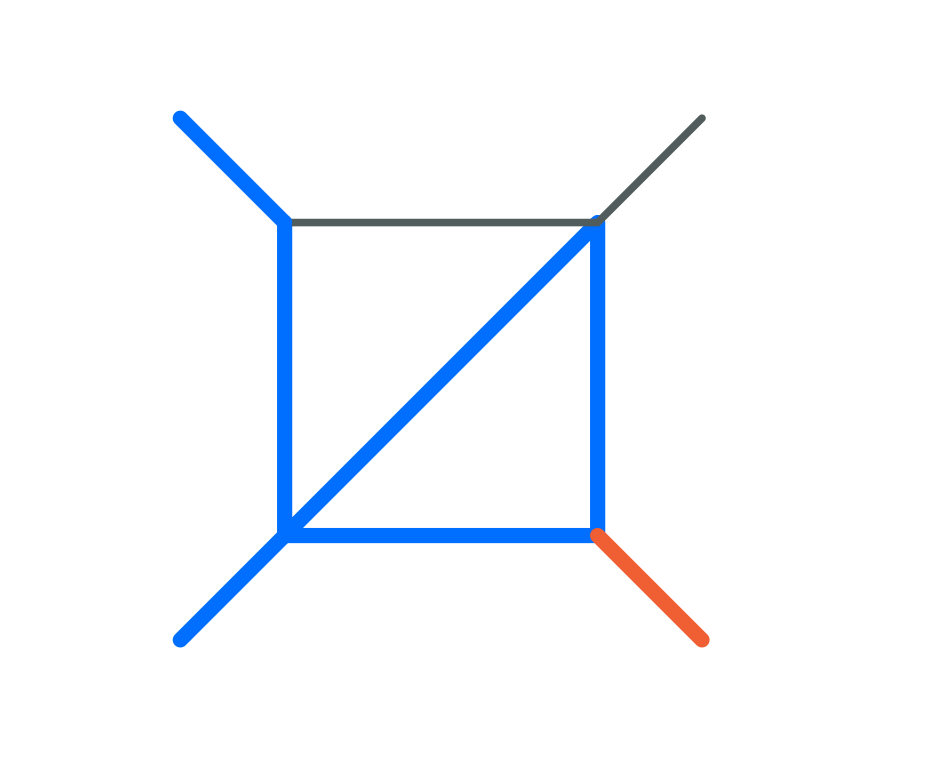}
    \end{subfigure}
    \begin{subfigure}{0.5\textwidth}
    \centering
    \includegraphics[width=0.9\linewidth]{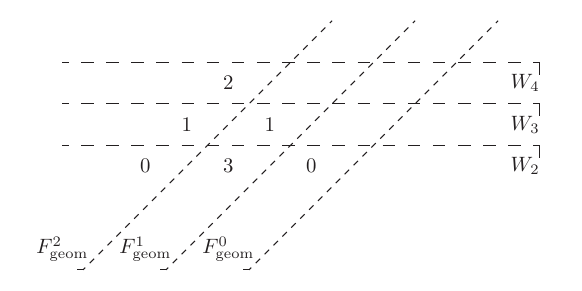}
    \end{subfigure}
    \caption{A kite diagram. Blue lines represent propagators and external momenta with the mass $m^2$, while the orange line represent the external momenta with the mass $q^2$. Thinner black lines are massless. The top sector has seven master integrals that decompose with respect to $\Fgeom^{\bullet}$ and $W_{\bullet}$ filtration as shown on the right figure.} 
    \label{fig:pl3}
\end{figure}

We are considering a kite diagram (see Fig.~\ref{fig:pl3}) with the following kinematics:
\begin{align}
    p_1^2&=p_2^2=m^2, \qquad p_3^2=0, \qquad p_4^2=q^2 \notag \\
    s&=(p_1+p_2)^2, \quad t=(p_2+p_3)^2,\quad u=(p_1+p_3)^2
\end{align}
The inverse propagators of the full family are defined as:
\begin{align}
    \sigma_1&=(k1_p1)^2, \quad \sigma_2=k_1^2-m^2, \quad \sigma_3=(k_1+p_2)^2, \quad \sigma_4=(k_2-p_1)^2-m^2 \notag \\
    \sigma_5&=(k_2+p_2) - m^2, \quad
    \sigma_6=(k_2+p_2+p_3)^2-m^2, \quad \sigma_7=(k_1-k_2)^2-m^2, \notag \\
    \sigma_8&=(k_1+p_2+p_3)^2, \quad \sigma_9=k_2^2.
\end{align}
We are interested in the kite sector whose inverse propagators are $\{\sigma_2, \sigma_3, \sigma_4,\sigma_6, \sigma_7 \}$, and we use two ISPs ($z_1=\sigma_5$ and $z_2=\sigma_9$) in order to compute a loop-by-loop Baikov representation. On the maximal cut we get:
\begin{equation}
    e^{2 \eps \gamma_E} \int_{\mathcal{C}_{\text{maxcut}}}\dfrac{d^Dk_1}{i \pi^{2-\eps}}\dfrac{d^Dk_2}{i \pi^{2-\eps}} \dfrac{1}{\sigma_2\sigma_3\sigma_4\sigma_6\sigma_7}=C_B \int \dfrac{dz_1}{2\pi i}\dfrac{dz_2}{2\pi i} p_1^{-2 \eps } {p_2}^{-\frac{1}{2} +\eps} {p_3}^{ -\frac{1}{2} -\eps},
\end{equation}
where 
\begin{align}
    \label{eq:poly3}
    C_B &= -\dfrac{(-1)^{-\eps} 16 e^{2 \eps \gamma_E} \pi^3}{\Gamma\left(1-2\eps \right)} \left(m^2 \right)^{-\eps} \left(-m^4 s+m^2 \left(-q^4+q^2 s+2 s t\right)-s t \left(-q^2+s+t\right)\right)^{\eps}, \notag \\
    p_1&=z_1, \notag  \\
    p_2&= \left(4 m^2 z_1-\left(z_1-z_2\right){}^2\right),  \\
    p_3&= z_1^2 \left(m^4-2 m^2 \left(q^2+t\right)+\left(q^2-t\right)^2\right) + z_2^2\left(q^2-s\right)^2 \notag\\
    & -2 z_1 z_2 \left(m^2 \left(q^2+s\right)+q^4-q^2 (s+t)-s t\right) + 2 z_1 \left(m^2-t\right) \left(m^2 \left(2 q^2+s\right)+s \left(t-q^2\right)\right) \notag \\
    & -2 z_2 \left(q^2-s\right) \left(m^2 \left(2 q^2-s\right)-s t\right) -s(4 m^2 - s) (m^2 - t)^2. \notag
\end{align}
The homogenous twist is 
\begin{equation}
    U= z_0^{2 \eps}P_1^{-2 \eps } {P_2}^{-\frac{1}{2}+\eps} {P_3}^{-\frac{1}{2}-\eps},
\end{equation}
where polynomials $P_1,\; P_2,\; P_3$ are homogenous versions of polynomials in equation~\eqref{eq:poly3}. The twist is of the degree $-2$. Therefore, differential forms have to be of a degree $-1$. There are seven master integrals in this sector. 

Polynomials $P_0$ and $P_1$ are even, hence we start the analysis by localising on them. Let us first look at the $P_0=0$ localisation. We may take two consecutive non-zero residue at points
\begin{equation}
    \left[ 0:0:1 \right], \quad \left[ 0:1:1 \right].
\end{equation}
The differential form $\Psi_{1100}[z_2]$ has residue at both points, while $\Psi_{1000}[1]$ has a residue only at the second point. We assign to these forms $a=-4$. Running the algorithm at weight 3 tells us that these two forms are independent and that the basis for $\langle P_0 \rangle$ localisation is 
\begin{equation}
    \{\Psi_{1000}[1],\; \Psi_{1100}[z_2] \}.
\end{equation}
Both of these forms have $w=4$ and $o=2$. 

We proceed with the localisation on the remaining even polynomial $ P_1=0$. The form $\Psi_{1100}[z_2]$ has two residues at the point 
\begin{equation}
    \left[ 0:1:1 \right],
\end{equation}
and we assign $a=-4$ to it. In addition, this form has $w=4$ and $o=2$. We then run the reduction at weight 3 and find the remaining 2 candidates
\begin{equation}
    \{\Psi_{0100}[1],\; \Psi_{0110}\left[ 2 m^2 (-m^2 + t) z_0 + (m^2 + t) z_1 + (m^2 - t) z_2\right] \}.
\end{equation}
Both of these forms have $w=3$, while the first has $o=1$ and the second one has $o=2$. We assign $a=-3$ to these forms before merging the results of the two localisations (see Ref.~\cite{Bree:2025tug} for details on the merge procedure). 

After the merge procedure, we are left with four independent master integrands
\begin{equation}
    \lbrace \Psi_{1000}[1],\; \Psi_{0100}[1],\; \Psi_{1100}[z_2] ,\; \Psi_{0110}\left[2 m^2 (-m^2 + t) z_0 + (m^2 + t) z_1 + (m^2 - t) z_2\right] \rbrace. 
\end{equation}
Running the Laporta algorithm on the full system, we find the remaining 3 master integrands
\begin{equation}
    \lbrace \Psi_{0001}[\mathcal{N}_1],\; \Psi_{0001}[\mathcal{N}_2],\; \Psi_{0010}[\mathcal{N}_3]\rbrace, 
\end{equation}
where
\begin{align}
    \mathcal{N}_1&= -(m^2 - t) (2 m^2 q^2 - m^2 s - s t) z_0 + \bigl[-m^4 + (q^2 - 2 s - t) t + 
    m^2 (q^2 + 2 t) \bigr] z_1 + (q^2 - s) (m^2 - t) z_2 \notag \\
    \mathcal{N}_2 &=- (4 m^2 - s) s (m^2 - t) z_0 + \bigl[m^2 (2 q^2 + s) + s (-q^2 + t)\bigr] z_1 + 
 s (2 m^2 + q^2 - s - 2 t) z_2, \notag \\
    \mathcal{N}_3&=2 m^2 (m^2 - t) z_0 + (2 m^2 - s - 2 t) z_1 - s z_2.
\end{align}
These 3 master integrands have $w=2$ and $o=1$. 

The full basis of master integrands ordered by $|\mu|$ is 
\begin{align}
   &\left \lbrace \Psi_{0001}[\mathcal{N}_1], \; \Psi_{0001}[\mathcal{N}_2], \; \Psi_{0010}[\mathcal{N}_3] ,\; \Psi_{1000}[1], \; \Psi_{0100}[1], \right. \\
   &\left. \; \Psi_{011}\left[2 m^2 (-m^2 + t) z_0 + (m^2 + t) z_1 + (m^2 - t) z_2\right] , \; \Psi_{1100}[z_2] \right \rbrace,
\end{align}
which corresponds to the basis
\begin{align}
    &\left\lbrace \frac{1}{2} \eps^2 I_{011201100}, \; \frac{1}{2} \eps^2 I_{011102100}, \right. \notag \\
    &\left.\frac{1}{2} \eps^2 \left[I_{(-1)21101100} - I_{0211011(-1)0} +\eps \dfrac{m^2-t- s}{m^2} I_{011101100} - s I_{021101100}\right], \right. \notag \\
    &\left.-\eps^2 I_{011101200},\; 2 \eps^3 I_{011101100}, 
    - 2 \varepsilon^2 \left(
        2 m^2 I_{021101100} +2 \varepsilon I_{011101100} + 2 m^2 I_{011101200}
        \right), \right. \notag \\
        &\left. \eps I_{0111013(-1)0} \right\rbrace 
\end{align}
on the Feynman integral side.

The resulting differential equation is in the $F^{\bullet}$-compatible form with the following $\eps$-dependence of the connection matrices:
\begin{equation}
    \left(
\begin{array}{ccccc|cc}
 \{0,1\} & \{0,1\} & \{1\} & \{0,1\} & \{1\} & \{1\} & \{1\} \\
 \{0,1\} & \{0,1\} & \{1\} & \{0,1\} & \{1\} & \{1\} & \{1\} \\
 \{0,1\} & \{0,1\} & \{1\} & \{0,1\} & \{0,1\} & \{1\} & \{1\} \\
 \{1\} & \{1\} & \{1\} & \{0,1\} & \{1\} & \{1\} & \{1\} \\
 \{1\} & \{1\} & \{1\} & \{1\} & \{0,1\} & \{1\} & -  \\
 \hline
 \{1\} & \{1\} & \{1\} & \{0,1\} & \{0,1\} & \{1\} & \{1\} \\
 \{0,1\} & \{0,1\} & \{0,1\} & \{-1,0,1\} & \{0,1\} & \{0,1\} & \{0,1\} \\
\end{array}
\right).
\end{equation}
Here $\{-1,0,1\}$ means that the coefficients of $\eps^{-1},\; \eps^0$ and $\eps^1$ are non-vanishing rational functions of kinematic variables.

Generally, the entries of connection matrices are at most be rational functions of $\eps$ when using one of the standard orderings in the Laporta algorithm. In this example, using one of the standard orderings in the Laporta algorithm, gives us a basis
\begin{align}
   \left\{ I_{011101100},\; I_{021101100},\; 
  I_{012101100},\; I_{011201100},\; 
  I_{011102100},\; I_{011101200}, \;
 I_{031101100} \right\},
\end{align}
whose differential equations have connection matrices with $\eps$-dependence as
\begin{equation}
    \left(
\begin{array}{ccccccc}
 \{0,1\} & \{-1,0\} & \{-1,0\} & \{0\} & \{0\} & \{-1,0\} & \{-1\} \\
 \{2\} & \{0,1\} & \{0,1\} & \{1\} & \{1\} & \{0,1\} & \{0\} \\
 \{2\} & \{0,1\} & \{0,1\} & \{1\} & \{1\} & \{0,1\} & \{0\} \\
 \{2\} & \{0,1\} & \{0,1\} & \{0,1\} & \{0,1\} & \{0,1\} & \{0\} \\
 \{2\} & \{0,1\} & \{0,1\} & \{0,1\} & \{0,1\} & \{0,1\} & \{0\} \\
 \{2\} & \{0,1\} & \{0,1\} & \{1\} & \{1\} & \{0,1\} & \{0\} \\
 \{2,3\} & \{0,1,2\} & \{0,1,2\} & \{1,2\} & \{1,2\} & \{0,1,2\} & \{0,1\} \\
\end{array}
\right).
\end{equation}
Even though, we only get monomials in $\eps$ when using one of the standard orderings, we still benefit from the improved ordering. The first benefit is that the size of the differential equations is roughly a half of the standard one. The second benefit is that we can systematically remove unwanted terms in $\eps$ until we reach an $\eps$-factorised form.

\section{Conclusions}
In this talk, we reviewed a recently proposed two-step algorithm for finding $\eps$-factorised differential equations for Feynman integrals. In the first step, we use a new ordering relation in the Laporta algorithm inspired by geometry. We observe that the resulting basis satisfies differential equations compatible with a filtration. This step involves only rational functions in kinematic variables which benefits semi-numerical approaches~\cite{Hidding:2020ytt,Armadillo:2022ugh,Prisco:2025wqs, PetitRosas:2025xhm}. Additionally, we observe a reduction in size of differential equations since our approach avoids spurious polynomials in the denominators. In the second step, we may systematically remove unwanted terms in $\eps$ starting from a differential equation compatible with a filtration.

\end{document}